\documentclass[aps,pre,twocolumn,showpacs,groupedaddress]{revtex4-1}

\usepackage{graphicx}
\usepackage{amssymb}
\usepackage{subeqn}

\begin{document}

\title{Experimental velocity fields and forces\\for a cylinder penetrating into a granular medium}

\author{A. Seguin$^1$, Y. Bertho$^1$, F. Martinez$^1$, J. Crassous$^2$ and P. Gondret$^1$}

\affiliation{$^1$Univ Paris-Sud, Univ Paris 6, CNRS, Lab FAST, B\^at.~502, Campus Univ, F-91405 Orsay, France}
\affiliation{$^2$Univ Rennes 1, Institut de Physique de Rennes (UMR UR1-CNRS 6251), B\^{a}t.~11A, Campus de Beaulieu, F-35042 Rennes, France}

\begin{abstract}
We present here a detailed granular flow characterization together with force measurements for the quasi-bidimensional situation of a horizontal cylinder penetrating vertically at a constant velocity in dry granular matter between two parallel glass walls. In the velocity range studied here, the drag force on the cylinder does not depend on the velocity $V_0$ and is mainly proportional to the cylinder diameter $d$. While the force on the cylinder increases with its penetration depth, the granular velocity profile around the cylinder is found to be stationary with fluctuations around a mean value leading to the granular temperature profile. Both mean velocity profile and temperature profile exhibit strong localization near the cylinder. The mean flow perturbation induced by the cylinder decreases exponentially away from the cylinder on a characteristic length $\lambda$ that is mainly governed by the cylinder diameter for a large enough cylinder/grain size ratio $d/d_g$: $\lambda \sim d/4 + 2 d_g$. The granular temperature exhibits a constant plateau value $T_0$ in a thin layer close to the cylinder of extension $\delta_{T_0} \sim \lambda/2$ and decays exponentially far away with a characteristic length $\lambda_T$ of a few grain diameters ($\lambda_T \sim 3 d_g$). The granular temperature plateau $T_0$ that scales as $V_0^2d_g/d$ is created by the flow itself from the balance between the ``granular heat" production by the shear rate $V_0/\lambda$ over $\delta_{T_0}$ close to the cylinder and the granular dissipation far away.
\end{abstract}


\maketitle

\section{Introduction}
The characterization of forces on moving objects in granular matter is important in fields ranging from fluid mechanics to geophysics and biophysics with, for instance, the practical situations of meteoritic impacts on planets or asteroids \cite{Melosh89} and of the motion of living organisms in sand \cite{Maladen09}. As a matter of fact, a better understanding of meteoritic impacts requires a better knowledge of the forces experienced by impactors, and this gave rise to numerous physical studies at the laboratory scale (\cite{Katsuragi07,Goldman08} and references herein) or by numerical simulations \cite{Bourrier08,Seguin09}. Most of these studies have identified mainly two terms in the forces experienced by an impactor: One term proportional to the square of the velocity and independent of the depth during the initial stage of the penetration (at high velocity) and one term proportional to the penetration depth and independent of the velocity at the end of the penetration (at low velocity) \cite{Katsuragi07,Goldman08,Seguin09}. The stopping time has been shown to display a non intuitive behavior, with a smaller value for larger impact velocity \cite{Katsuragi07,Goldman08,Seguin09}. These behaviors for the penetration of the impactor may depend on the packing fraction of the grains \cite{Umbanhowar10}; the packing fraction also determines the grain ejection, which can give rise to a spectacular upwards vertical jet for a fluidized packing \cite{Thoroddsen01,Mikkelsen02} or to an opening corona for a dense packing \cite{Deboeuf09,Marston12}, quite similar to what is observed for splashes in water of droplets or solid spheres \cite{Yarin06}. Moreover, the motion of certain living organisms in grains, such as sand snakes, is interesting to understand in order to, for example, create artificial robotics able to move in grains \cite{Maladen09}. In these last phenomena, the coupling between lift and drag forces is important \cite{Ding11}. All these physical phenomena require a better knowledge of the flow characterization around moving objects in grains and the associated forces. In this work, we shall focus on low velocity intrusions.

Many studies have been performed on the drag force experienced by objects in relative vertical or horizontal motion with grains, but fewer studies have been done concerning grain flow characterization. In dense granular matter, the drag force measured in velocity-controlled experiments has been found not to depend on the velocity at low velocities, and to be proportional to the surface area of the object and roughly to its depth \cite{Albert99}. As in classical fluids, the corresponding drag force has been shown to depend on the exact shape of the object \cite{Albert01}. This force may depend on the packing fraction \cite{Schroter07}, on the vibration of the grains \cite{Caballero09}, or on possible dynamical air effects \cite{Clement11}. Amarouch\`ene {\it et al.} reported a kinematic study of the flow around a thin disk placed in a Hele-Shaw cell and submitted to a vertical granular chute flow \cite{Amarouchene01}. The perturbed streamlines were observed to be located in a parabolic shape near the object with a triangular region of non flowing or slowly creeping grains, clearly due to the stabilizing friction force from the two close walls \cite{Courrech03,Taberlet03}. The measured velocity profile exhibited a non linear but monotonic variation away from the object. Chehata {\it et al.} performed the same kind of experiment for a cylinder in a much larger channel, and they measured at the same time the velocity field using a particle image velocimetry (PIV) technique and the drag force experienced by the cylinder in the grain chute flow \cite{Chehata03}. They reported measurements of the flow vorticity and velocity fluctuations that showed shear localization near the cylinder, and they noticed that the drag force is independent of the grain velocity and proportional to the cylinder diameter. In a horizontal bidimensional experiment of vibrated disks at different packing fraction close to jamming, the measured velocity field around an intruder dragged at a constant force also exhibited shear localization \cite{Candelier09}. In a horizontal quasi bidimensional experiment in which a thin but large cylinder is pulled relative to a dense packing of millimetric alumina beads, Takehara {\it et al.} \cite{Takehara10} investigated motion at high velocities for which the force is shown to increase quadratically with velocity, and they observed high grain fluctuation motions and velocity perturbation in a narrow zone around the cylinder that increases with the cylinder diameter. In the case of a dilute flow, the drag force is also observed to be proportional to the square of the velocity, and proportional to the cylinder diameter and to the effective density of the fluid as in classical hydrodynamics with a drag coefficient that increases from about 1 at low Kn to about 2-2.5 at high Kn \cite{Wassgren03,Boudet10}, where the Knudsen number Kn compares the mean free path in the gas to the object size.

In a previous work \cite{Seguin11}, we reported key results for both the experimental measurements and a hydrodynamic modeling of the canonical problem of the flow around a circular cylinder and the associated drag force in the granular case. In the present paper, we report the detailed experimental study of the mean grain velocity profile and fluctuations around a horizontal cylinder penetrating vertically in granular matter together with drag force measurements. In Sec.~\ref{sec:setup}, we present the experimental set-up. Experimental results on the profiles of both the mean velocity and the velocity fluctuations are presented in Sec.~\ref{sec:velo_temp}, whereas Sec.~\ref{sec:force} concerns force measurements. The results are discussed in Sec.~\ref{sec:discussion} before the conclusion.

\section{Experimental setup}
\label{sec:setup}
\begin{figure}[t!]
\centerline{\includegraphics[width=8cm]{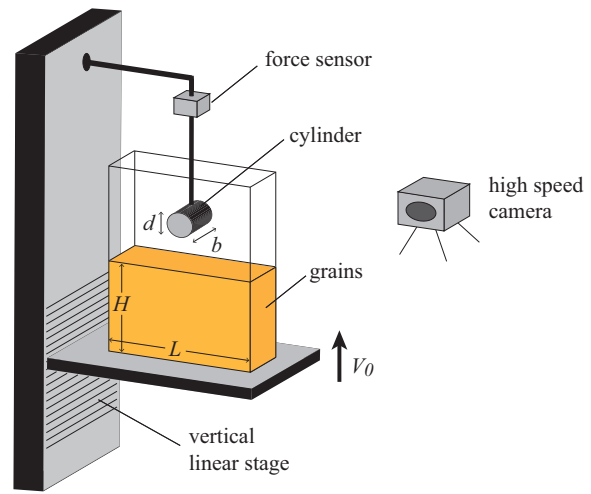}}
\caption{(Color online) Sketch of the experimental setup.}
\label{Fig01}
\end{figure}
\begin{figure}
\centerline{\includegraphics[width=\linewidth]{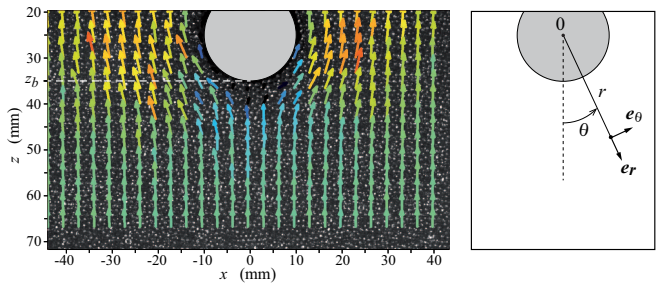}}
\caption{(Color online) Left: Typical instantaneous velocity field obtained by PIV measurements for a cylinder of diameter $d=20$\,mm penetrating in a packing of glass beads of diameter $d_g=1$\,mm, at the velocity $V_0=50$\,mm s$^{-1}$. Right: Sketch of the cylindrical coordinates and used notations.}
\label{Fig02}
\end{figure}

The experiments consist in the vertical penetration of a horizontal cylinder at a given velocity $V_0$ into a packing of grains. The grains are rather monodisperse sieved glass beads of diameter $d_g$ ranging from 0.5 to 4\,mm with a relative dispersion $\Delta d_g/d_g$ of about 10\% around the mean value, and density $\rho_g=2.5\,10^3$\,kg m$^{-3}$, contained in a rectangular box of length $L = $ 0.2\,m, height $H = $ 0.1\,m and width $b=40$\,mm (Fig.~\ref{Fig01}). The granular medium is prepared by gently stirring the grains with a thin rod, and the surface is then flattened using a straightedge. We have checked that this preparation leads to reproducible results with only small variations. The solid volume fraction is $\Phi \simeq 0.62$ characteristic of a dense granular packing, and the density of the granular medium is thus $\rho = \rho_g \Phi \simeq 1.5\ 10^3$\,kg m$^{-3}$. The cell containing the granular packing can be moved up along a vertical translation guide at a given velocity $V_0$ ranging up to 50\,mm s$^{-1}$ by a step by step motor (Mavilor BLS-55). A steel cylinder of length $b$ and of diameter $d$ ranging from 10 to 40\,mm is first maintained above the grains in between two vertical glass walls. The length $b$ of the cylinder is manufactured about 0.2\,mm smaller than the width $b$ of the cell, and the cell is free to move along the axial direction of the cylinder. In this way, the cylinder has only minimal mechanical contact and thus solid friction with the walls during its vertical displacement, and no grain can fit between the cylinder and the walls and jam the system. As the cell moves upwards, the cylinder penetrates gradually at a constant velocity in the granular medium. The entire dynamics of the grains during the penetration is extracted by recording their motion at the front glass wall using a high-speed video camera that can take up to 1000 images per second in the full resolution $1024\times 1024$ pixels. The grains are lighted from the front and a black curtain is put behind the cell so that the grains appear in a white on black background with a good contrast. The images are taken at a sampling rate $f$ adjusted on the velocity $V_0$ to have $f\geq V_0/d_g$ ({\it e.g.} $f=50$\,Hz for $V_0=10$\,mm s$^{-1}$ and $d_g=1$\,mm) so that the largest grain displacement between two successive images is smaller than one grain diameter. The successive images are then analyzed by Particle Image Velocimetry (PIV) software (Davis, LaVision) to get the velocity fields of the grains as shown in Fig.~\ref{Fig02}. The size of the final interrogation windows used in the correlation technique is set typically to one grain diameter to have the best resolution considering the discrete nature of the medium. The spatial resolution of the obtained velocity field is thus one grain diameter. As the cylinder and video camera are fixed in the laboratory frame of reference and the grains move up, the velocity field of the grains is measured in the frame of reference of the cylinder. The grain velocity far from the cylinder thus corresponds to the undisturbed upward imposed velocity $V_0$, whereas the velocity is disturbed close to the cylinder. Considering the discrete nature of the granular medium and the finite size of the PIV correlation window corresponding to one grain diameter, the PIV calculation is distorted at a distance from the cylinder surface smaller than half a grain diameter so that velocity measurements will be reported only outside this zone for radial distances $r$ from the cylinder such as $r > (d+d_g)/2$. With a force sensor (FGP Instrumentation FN3030) of range 50\,N, we also measure in the mean time the drag force experienced by the cylinder during its penetration in the granular packing at the sampling rate 2\,kHz. The cylinder is related to the force sensor by a vertical thin rod of diameter 3\,mm and length 10\,cm. Without grains in the cell, we checked that the friction force of the cylinder on the glass walls is totally negligible. With grains in the cell, we made sure that no grain was fitting into the small spacing between the cylinder sides and the walls and was distorting the force measurement during the cylinder penetration. In the following we present velocity and force measurements that have been done for a cylinder/grain size ratio in the range $10 \leqslant d/d_g \leqslant 80$ and for a cylinder velocity $0.5 \leqslant V_0 \leqslant 50$\,mm s$^{-1}$ remaining smaller than $(gd_g)^{1/2} \geqslant 70$\,mm s$^{-1}$. In this velocity range, the flow regime is expected to be quasi-static. The box length $L $ was chosen large enough ($L/d > 5$) to not play any significant role in the measurements \cite{Seguin08}. We also restrict ourselves to penetration depth values $z_b$ not too close to $H$ ($z_b \lesssim H-d$) so that there is no significant role played by the bottom wall \cite{Stone04,Seguin08}, and $H$ will thus not be a relevant parameter for the present results.

\section{Mean velocity profiles and velocity fluctuations}
\label{sec:velo_temp}
\subsection{Mean velocity}
\label{subsec:meanvelocity}

\begin{figure}
\centerline{\includegraphics[width=8cm]{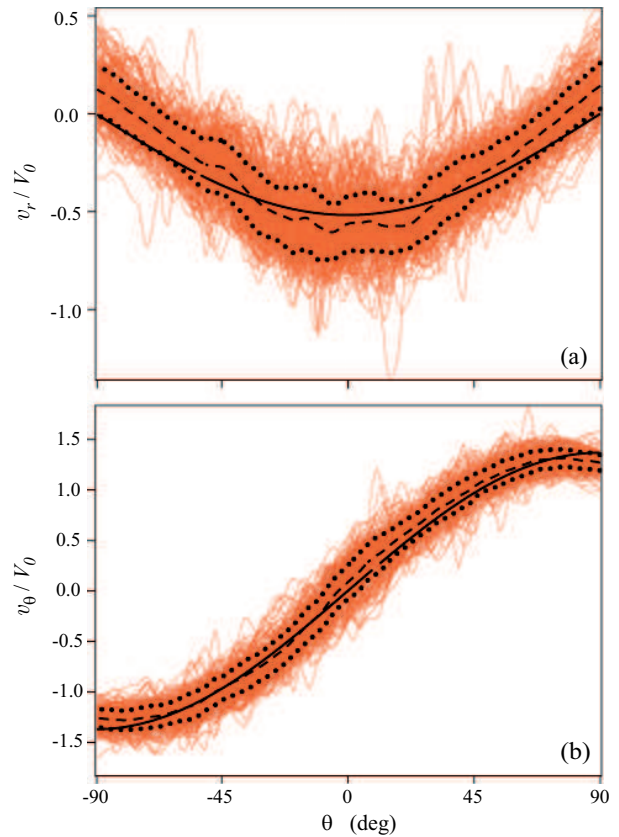}}
\caption{(Color online) All 230 successive instantaneous (a)~radial and (b)~azimuthal velocity profiles normalized by the penetration velocity $V_0$ taken at the radial location $r =15$\,mm at regular intervals ($\Delta t=20$\,ms) during the penetration interval $20 \leq z_b\leq 70$\,mm ($d=20$\,mm, $d_g=1$\,mm, $V_0=10$\,mm s$^{-1}$). (--~--)~Time average value and ($\cdot\cdot\cdot$) corresponding standard deviation. (---)~Cosine and sine fits of data by Eq.~(\ref{eq2}) with $A_r \simeq 0.5$ and $A_{\theta}\simeq 1.4$.}
\label{Fig03}
\end{figure}

The instantaneous velocity field obtained at each time $t$ by the PIV technique is decomposed within the cylindrical coordinates:
\begin{equation}
\mathbf{v}(r,\theta,t)=v_r(r,\theta,t)\ \mathbf{e_r} + v_\theta(r,\theta,t)\ \mathbf{e_\theta},
\label{eq1}
\end{equation}
where $r$ is the distance from the cylinder center and $\theta$ is the angle from the direction of motion ($\theta > 0$ anticlockwise) (Fig.~\ref{Fig02}). We have checked that the granular flow is bidimensional: No transverse flow along the $y$ direction perpendicular to the glass side wall can be seen either from the side (no grain appears at the glass wall or disappears from the glass wall during the cylinder penetration) or from the top (no transverse roll motion can be seen at the free surface). In addition, the results presented in the following will be for $-\pi/2<\theta<\pi/2$, \emph{i.e.} for the upstream grain flow, and we have checked that the downstream flow is very similar, with thus an upstream-downstream symmetrical flow field characteristic of low Reynolds numbers flows. Figure~\ref{Fig03} shows the successive instantaneous (a)~radial velocity component $v_r(r,\theta,t)$ and (b)~azimuthal velocity component $v_\theta(r,\theta,t)$ taken at a given radial location $r$ (here $r - d/2=5$\,mm), as a function of $\theta$ and normalized by the penetration velocity $V_0$. These velocity profiles are quite noisy as they are not averaged at all but they do not show any systematic time evolution during the penetration. This observation holds regardless of the radial position and the run. The velocity profile around the penetrating cylinder can thus be considered as stationary and is now time-averaged during each penetration run. The corresponding time average $\mathbf{v}(r,\theta)=$ $\langle\mathbf{v}(r,\theta,t)\rangle_t$ of the instantaneous velocity profiles appears as a thick dashed line in Fig.~\ref{Fig03} and the corresponding standard deviation is indicated by the surrounding two thick dotted lines. The radial velocity $v_r(r,\theta)$ is maximum near the direction of motion ($\theta \simeq 0$) and vanishes around the equatorial plane ($\theta \simeq \pm \pi/2$), and the opposite is true for the azimuthal velocity $v_\theta(r,\theta)$, which is maximal near the equator and vanishes near the axis of motion. This suggests the usual simple $\theta$ dependence for the two cylindrical velocity components:
\begin{equation}
\mathbf{v}(r,\theta)/V_0=-A_r(r)\cos\theta\ \mathbf{e_r} + A_\theta(r)\sin\theta\ \mathbf{e_\theta}.
\label{eq2}
\end{equation}

\begin{figure}
\centerline{\includegraphics[width=8cm]{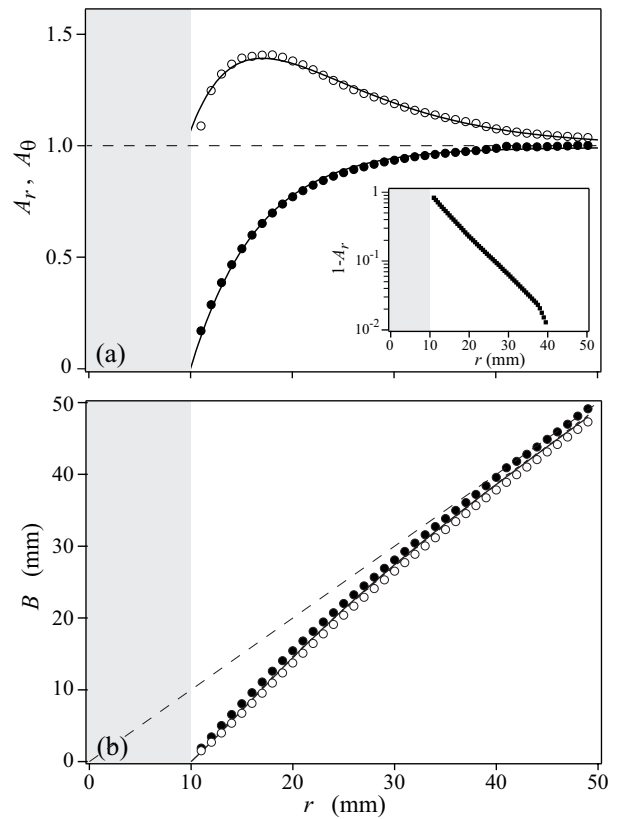}}
\caption{(a)~Radial dependence of the radial and azimuthal velocity components ($\bullet$)~$A_r$ and ($\circ$)~$A_\theta$, for $V_0=10$\,mm s$^{-1}$, $d=20$\,mm and $d_g=1$\,mm. (---) Best fits by Eqs.~(\ref{eq3}) with $\lambda \simeq 6.4$\,mm and $\lambda_s \simeq 9.7$\,mm for $A_r(r)$ data, and $\lambda \simeq 7.6$\,mm and $\lambda_s \simeq 8.1$\,mm for $A_\theta(r)$ data. Inset: 1-$A_r(r)$ data in log-lin plot. (b)~Radial dependence of the stream function, $B(r)$, obtained either from ($\bullet$) $A_r$ or ($\circ$) $A_\theta$ data, and (---) best fit from Eq.~(\ref{eq5}) with $\lambda \simeq 7.1$\,mm and $\lambda_s \simeq 9.2$\,mm. The grey zone corresponds to the cylinder interior ($r\leq 10$\,mm). The dashed lines stand for (a)~$A_{r,\theta}=1$ and (b)~$B(r)=r$ which corresponds to an undisturbed velocity field.}
\label{Fig04}
\end{figure}

The fits of experimental $v_r$ and $v_\theta$ data by cosine and sine functions are shown as thick solid lines in Fig.~\ref{Fig03} and are rather close to the averaged profiles. Doing such an analysis for all the radial locations $r$ allows us to obtain the radial functions $A_r(r)$ and $A_\theta(r)$, which prescribes the radial variation of $v_r$ and $v_\theta$. Such functions are plotted in Fig.~\ref{Fig04}(a) and show a strong flow localization around the penetrating cylinder.
As a matter of fact, $A_r(r)$ and $A_\theta(r)$ tend toward the value 1, corresponding to an undisturbed velocity field as soon as $ r \gtrsim 40$\,mm and thus $r \gtrsim 2d$. The radial function of the azimuthal velocity $A_\theta(r)$ exhibits an overshooting maximum ($A_\theta > 1$) as a consequence of the incompressibility for the present two-dimensional (2D) flow. This maximum is located very near the cylinder at $r \simeq 17$\,mm $\lesssim d$ and is followed by an inflection point. The radial function $A_r(r)$ of the radial velocity increases from zero at the cylinder surface up to the asymptotic value 1 at large $r$, with an exponential law as shown by the straight behavior of $1 -A_r(r)$ in the semi-logarithmic plot in the inset of Fig.~\ref{Fig04}(a). With this finding, assuming an incompressible flow with $div\ \mathbf{v} = (1/r) \partial(rv_r)/\partial r + (1/r) \partial v_\theta / \partial \theta = 0$, and taking into account the slip for the tangential velocity at the cylinder surface $v_{\theta}(r=d/2) \neq 0$, we adopt thus the following expressions for fitting $A_r(r)$ and $A_\theta(r)$ data:
\begin{subeqnarray}\label{eq3}
  A_r(r) & = &
    \frac{r- d/2 + \lambda_s}{r} \left [ 1-\exp \left (-\frac{r-d/2}{\lambda}\right )\right ],\label{eq3a}\\
  A_\theta(r) & = &
    1+\frac{r- d/2 + \lambda_s - \lambda}{\lambda}\exp\left (-\frac{r-d/2}{\lambda}\right ),
  \label{eq3b}
\end{subeqnarray}
where $\lambda$ is the characteristic length of velocity radial variation, and the length $\lambda_s$ characterizes the velocity slip tangential to the cylinder surface. As a matter of fact, $A_\theta(r=d/2)=\lambda_s/\lambda$ so that there is some slip for $\lambda_s \neq 0$. The slipping length $l_s$ defined as $(\partial v_\theta/\partial r)(r=d/2) = v_\theta(r=d/2)/l_s$ is related to $\lambda_s$ by $l_s = \lambda_s/(2-\lambda_s/\lambda)$. As experimentally $A_\theta(r=d/2)=\lambda_s/\lambda \simeq 1$, we have $l_s \simeq \lambda_s$. Thus the length $\lambda_s$ corresponds roughly here to the slipping length $l_s$. The curves corresponding to Eqs.~(\ref{eq3}) are plotted as solid lines in Fig.~\ref{Fig04}(a) with the best-fitting values $\lambda \simeq 6.4$\,mm and $\lambda_s \simeq 9.7$\,mm for $A_r(r)$ data, and $\lambda \simeq 7.6$\,mm and $\lambda_s \simeq 8.1$\,mm for $A_\theta(r)$ data. The fitting curves are close to the data, and the $\lambda$ and $\lambda_s$ values found independently from $A_r(r)$ and $A_\theta(r)$ data are close, which means that the present dense granular flow is nearly incompressible. Note that as $\lambda_s$ values are close to half the cylinder diameter the prefactor $(r-d/2+\lambda_s)/r$ in Eq.~(\ref{eq3a}) is close to one, thus explaining the exponential behavior observed for $1-A_r$ [inset of Fig.~\ref{Fig04}(a)].

\begin{figure}
\centerline{\includegraphics[width=\linewidth]{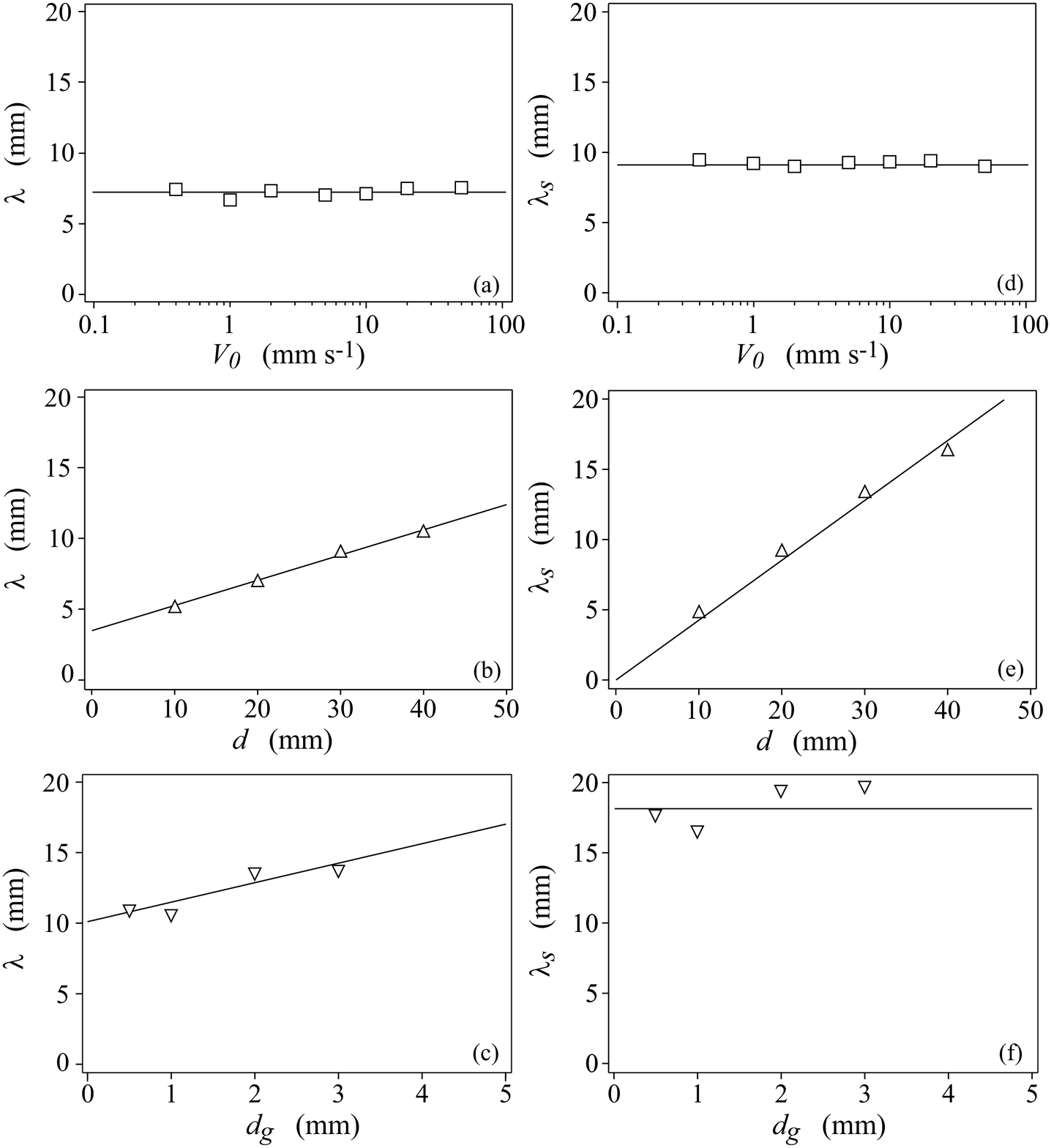}}
\caption{Variations of the characteristic length $\lambda$ and $\lambda_s$ of mean velocity profile with (a, d)~the velocity $V_0$ (for $d=20$\,mm and $d_g=1$\,mm), (b, e)~the cylinder diameter $d$ (for $V_0=5$\,mm s$^{-1}$ and $d_g=1$\,mm) and (c, f)~the grain size $d_g$ (for $V_0=5$\,mm s$^{-1}$ and $d=40$\,mm). The symbols are experimental data and the solid lines correspond to (a)~$\lambda =7$\,mm, (b)~$\lambda =3\mathrm{\,mm}+0.2d$, (c)~$\lambda =10\mathrm{\,mm} + 1.5d_g$, (d)~$\lambda_s=9$\,mm, (e)~$\lambda_s=0.4d$, and (f)~$\lambda_s=18$\,mm.}
\label{Fig05}
\end{figure}

As the flow can be considered to be bidimensional in the ($r$, $\theta$) plane, one can introduce the scalar stream function $\psi(r,\theta)$ related to the velocity field by $v_\theta = -\partial \psi/\partial r$ and $v_r = (1/r) \partial \psi / \partial \theta$. Considering the $\theta$ dependence for the velocity field of Eq.~(\ref{eq2}), the stream function can be written as
\begin{equation}
\psi(r,\theta)=-V_0\ B(r)\sin\theta,
\label{eq4}
\end{equation}
where the radial function $B(r)$ is related to the velocity radial functions $A_r(r)$ and $A_\theta(r)$ by $A_r=B/r$ and $A_\theta=dB/dr$. The $B(r)$ curves obtained either from $A_r(r)$ or $A_\theta(r)$ data of Fig.~\ref{Fig04}(a) are shown in Fig.~\ref{Fig04}(b). These curves are close, meaning again that the present granular flow is close to being truly incompressible. The average experimental $B(r)$ data can be fitted by the following empirical function deduced from Eqs.~(\ref{eq3}):
\begin{equation}
B(r)=(r- d/2 + \lambda_s)\left [ 1-\exp \left (-\frac{r-d/2}{\lambda}\right )\right ].
\label{eq5}
\end{equation}
For large radial distance $r$ away from the cylinder ($r \gtrsim 2d$), $B(r)$ is close to linear in $r$, which corresponds to the undisturbed flow. The interesting domain is thus the one close to the cylinder ($r \lesssim2d$) where $B(r)$ deviates significantly from $r$. The fit of the mean $B(r)$ data of Fig.~\ref{Fig04}(b) by the previous equation leads to $\lambda=7.1$\,mm and $\lambda_s=9.2$\,mm, which correspond to intermediate values between the $\lambda$ and $\lambda_s$ values deduced directly from $A_r(r)$ and $A_\theta(r)$ data. In the following, the presented $\lambda$ and $\lambda_s$ values have been obtained by the fit of $B(r)$ experimental data.

The $\lambda$ and $\lambda_s$ variations with the penetration velocity $V_0$, the cylinder diameter $d$, and the grain diameter $d_g$ are shown in Fig.~\ref{Fig05}. One can see that neither $\lambda$ nor $\lambda_s$ depend on the cylinder velocity $V_0$ [Figs.~\ref{Fig05}(a,d)] with the nearly constant values $\lambda = 7 \pm 1$ mm and $\lambda_s = 9\pm 1$ mm. The $\lambda_s$ values appear to be proportional to the cylinder diameter $d$ with the linear fit $\lambda_s = 0.4 d$ very close to the data [Fig.~\ref{Fig05}(e)], and no significative dependence upon the grain diameter $d_g$ can be seen in Fig.~\ref{Fig05}(f) with the nearly constant value $\lambda_s = 18 \pm 2$\,mm corresponding again to about $0.4d$. The $\lambda$ values increase linearly with the cylinder diameter $d$ with a slope around 0.2 and a significative non zero value around 3\,mm corresponding here to $3 d_g$ that may be extrapolated for vanishing $d$ [Fig.~\ref{Fig05}(b)]. The $\lambda$ variations with the grain diameter $d_g$ shown in Fig.~\ref{Fig05}(c) are small but may increase from about 10 to 15\,mm with an extrapolated value $d/4$ for vanishing $d_g$ and a slope around 1.5. In conclusion, these two characteristic lengths $\lambda$ and $\lambda_s$ do not depend on the penetration velocity $V_0$ and are mainly governed by the cylinder diameter $d$ with a possible weak effect of the grain size $d_g$ on $\lambda$ when the cylinder diameter is not much larger than the grain size.

\subsection{Velocity fluctuations}

\begin{figure}
\centerline{\includegraphics[width=\linewidth]{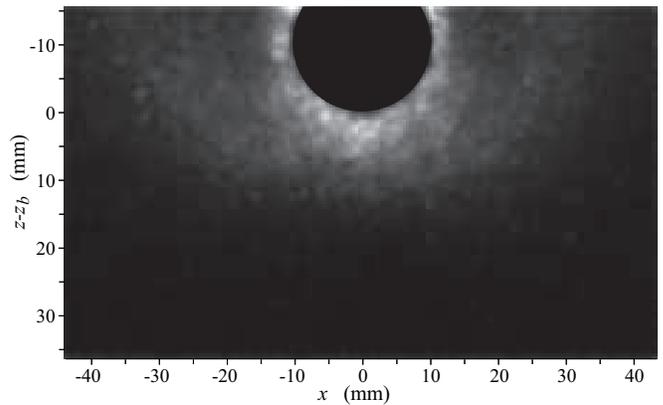}}
\caption{Temperature field $T(x,z)$ around a cylinder of diameter $d=20$\,mm penetrating with the velocity $V_0=10$\,mm s$^{-1}$ in a packing of grains of diameter $d_g=1$\,mm.}
\label{Fig06}
\end{figure}

\begin{figure}
\centerline{\includegraphics[width=8cm]{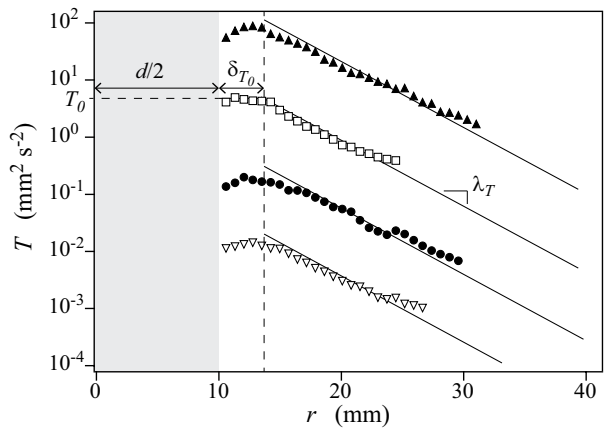}}
\caption{Radial dependence of the temperature profile $T(r)$ for $\theta =0$, with $d=20$\,mm, $d_g=1$\,mm and different velocities ($\triangledown$)~$V_0=0.4$\,mm s$^{-1}$, ($\bullet$)~$V_0=2$\,mm~s$^{-1}$, ($\square$)~$V_0=10$\,mm s$^{-1}$ and ($\blacktriangle$)~$V_0=50$\,mm s$^{-1}$. Guidelines for the eyes with (---)~the slope value $\lambda_T=3.7$\,mm and (- - -)~the plateau value $T_0$ and radial extension $\delta_{T_0} = 4$\,mm close to the cylinder surface at $r = d/2$.}
\label{Fig07}
\end{figure}

During each penetration run, we observe that the velocity field is stationary with a well defined time average value and some erratic time fluctuations. In the preceding section, we focused on the time average velocity field $\mathbf{v}(r,\theta)$, and we now present the time fluctuations whose amplitude can be quantified by the granular temperature defined as $T(r,\theta) = \langle\bigl(\mathbf{v}(r,\theta,t)-\mathbf{v}(r,\theta)\bigr)^2\rangle_t$.
A typical temperature field $T(r,\theta)$ around the cylinder is shown in Fig.~\ref{Fig06} in gray scale with black for the lowest value and white for the highest value. A zone of high temperature can be seen very near the cylinder with a radial extension of a few millimeters around the cylinder surface. The precise radial dependence of the temperature extracted along the $\theta = 0$ streamwise direction in front of the cylinder is plotted in Fig.~\ref{Fig07} for different cylinder velocities $V_0$. One can see in this semi logarithmic plot that all these different profiles have the same shape with about a plateau value denoted $T_0$ in a zone of extension $\delta_{T_0}$ close to the cylinder ($0 < r - d/2 < \delta_{T_0}$) followed by an exponential decrease $T \sim \exp(-r/\lambda_T)$ with the characteristic length $\lambda_T$ at larger $r$ ($r - d/2 > \delta_{T_0}$). The different curves at different velocities $V_0$ show that $T_0$ increases with $V_0$ whereas neither $\delta_{T_0}$ nor $\lambda_T$ varies significantly as the decreasing parts of $T(r)$ are about parallel: $\delta_{T_0} \simeq 4$\,mm and $\lambda_T \simeq 3.7$\,mm.

\begin{figure}
\centerline{\includegraphics[width=\linewidth]{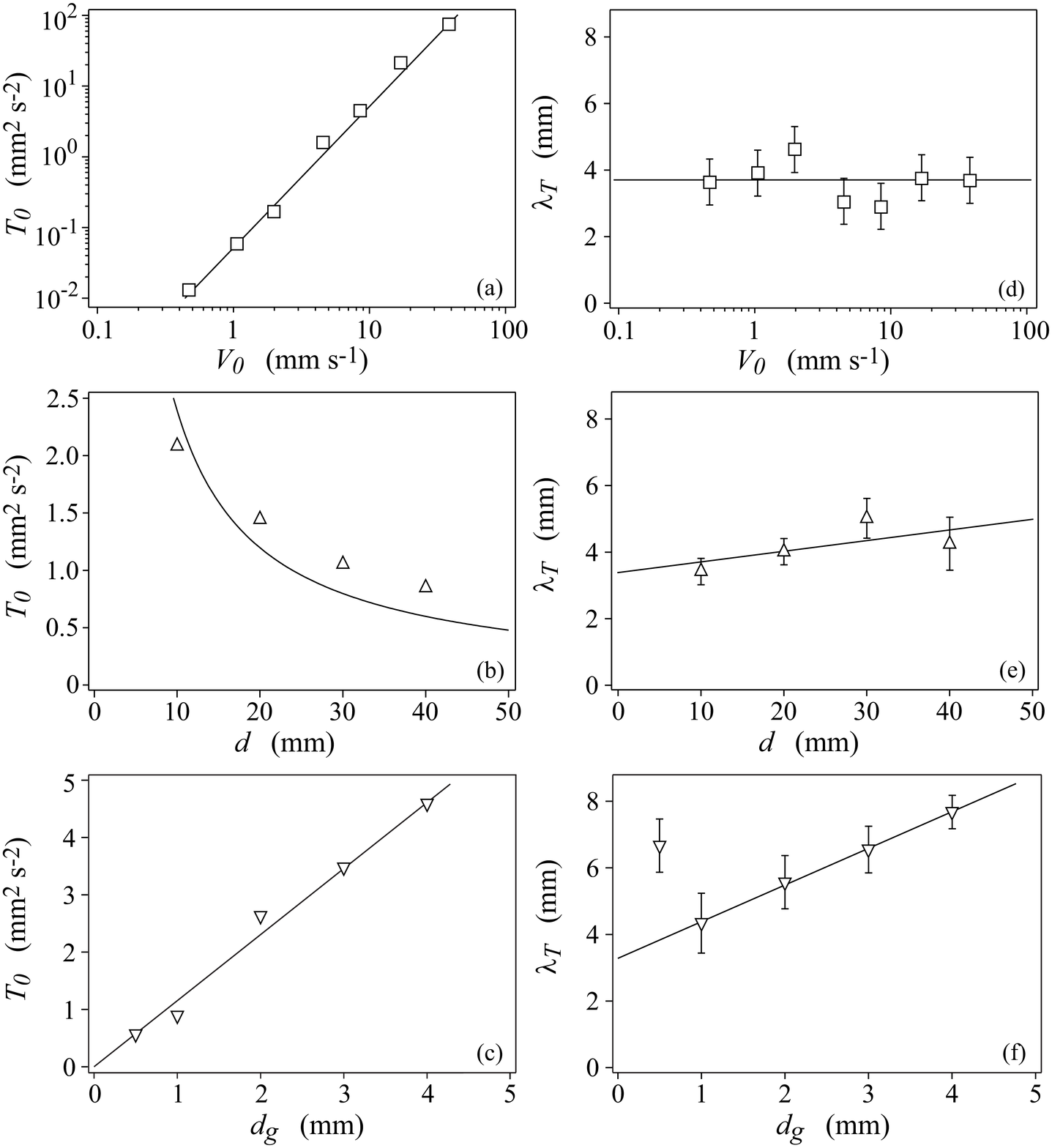}}
\caption{Temperature plateau $T_0$ and characteristic length $\lambda_T$ of radial temperature variation as a function of (a, d)~the velocity $V_0$ (for $d=20$\,mm and $d_g = 1$\,mm), (b, e)~the cylinder diameter $d$ (for $V_0 =5$\,mm s$^{-1}$ and $d_g=1$\,mm) and (c, f)~the grain size $d_g$ (for $V_0=5$\,mm s$^{-1}$ and $d=40$\,mm). The symbols are experimental data and the solid lines correspond to (a)~$T_0 = 0.05V_0^2$, (b)~$T_0\propto d^{-1}$, (c)~$T_0\propto d_g$, (d)~$\lambda_T=3.7$\,mm, (e)~$\lambda_T=3.5$\,mm$+0.03d$, and (f)~$\lambda_T=3.3$\,mm $+ 1.1d_g$.}
\label{Fig08}
\end{figure}

\begin{figure}
\centerline{\includegraphics[width=0.7\linewidth]{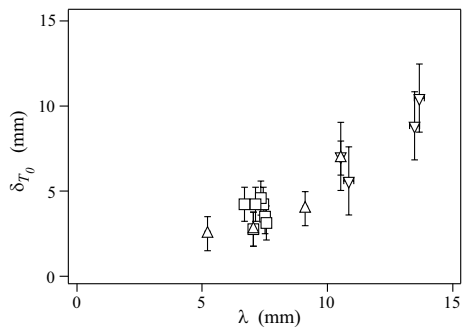}}
\caption{Radial extension $\delta_{T_0}$ of the temperature plateau as a function of the characteristic length $\lambda$ of the mean velocity profile (same symbols as in Fig.~\ref{Fig08}).}
\label{Fig09}
\end{figure}

The variations of $T_0$ and $\lambda_T$ with the velocity $V_0$, the cylinder diameter $d$, and the grain diameter $d_g$ are shown in Fig.~\ref{Fig08}. The log-log plot of $T_0$ with $V_0$ in Fig.~\ref{Fig08}(a) exhibits a clear data increase with a slope 2 leading to the natural scaling $T_0 \sim V_0^2$. Figures~\ref{Fig08}(b) and \ref{Fig08}(c) show that the typical temperature $T_0$ decreases with increasing cylinder size $d$ and increases with the grain size $d_g$. The $T_0$ dependence upon $d_g$ is linear and the $T_0$ dependence upon $d$ may be hyperbolic such that $T_0$ may finally scale as $V_0^2 (d_g/d)$, which will be supported by the plot of Fig.~\ref{Fig13}(a), as we shall see in the discussion (Sec. V). Concerning the characteristic length $\lambda_T$ of the exponential temperature decrease away from the cylinder, Fig.~\ref{Fig08}(d) shows that $\lambda_T$ does not depend on $V_0$ with the nearly constant value $\lambda_T=(3.7\pm 0.5)$\,mm. One can see in Fig.~\ref{Fig08}(e) that $\lambda_T$ does not depend significantly on the cylinder diameter $d$ with a value around $\lambda_T\simeq 4$\,mm corresponding thus to a few grain diameters. This is corroborated by Fig.~\ref{Fig08}(f), where $\lambda_T$ appears to increase linearly with the grain diameter $d_g$ if one ignores the singular point at the smallest grain size $d_g=0.5$\,mm. The dependence of the radial width of the temperature plateau $\delta_{T_0}$ on $V_0$, $d$, and $d_g$ is not shown independently, but Fig.~\ref{Fig09} displays a clear correlation of $\delta_{T_0}$ with the characteristic length $\lambda$ of mean velocity variations. This means thus that $\delta_{T_0}$ does not depend on $V_0$ and is mainly governed by the cylinder diameter $d$ with only a weak effect of the grain size $d_g$ in contrast with the $\lambda_T$ size dependence discussed above.

\ \\

\section{Force on the cylinder}
\label{sec:force}

The force on an object in relative motion with a granular material has been extensively studied these last few years by many authors in a lot of geometries. The drag force is observed to vary linearly with the surface area of the object and increases with the depth $z$, $F\sim z^\alpha$, with possible non linear effects ($\alpha\neq1$) depending on the direction of motion and geometry \cite{Albert99, Albert01, Chehata03, Stone04, Hill05, Peng09}. For a vertical rod in a horizontal granular flow, $\alpha$ is significantly larger than 1 ($\alpha\simeq 2$ \cite{Albert99, Reddy11}), which is linked to the linear increase of the immersed surface with $z$. For objects whose immersed area does not vary with the depth, the exponent $\alpha$ is closer to 1. For the vertical penetration of spheres, cubes, and horizontal rods, $\alpha\simeq 1.2$ was found \cite{Hill05}. For the vertical penetration of spheres and also of cylinders and cones with aspect ratio one and vertical axis, Ref.~\cite{Peng09} found that $\alpha\geq1$,with a $\alpha$ value depending on the geometry: $\alpha\simeq1$ for these cylinders, $\alpha\simeq1.3$ for spheres, and $\alpha\simeq1.8$ for these cones. This supralinear depth dependence of the force ($\alpha\geq1$) is observed at small depth but is followed by a sublinear dependence ($\alpha\leq1$) at large depth attributed to the counter force generated by the filled-in grains on the top of the object \cite{Peng09}. For a horizontal plate penetrating a granular medium, Ref.~\cite{Stone04} found a linear increase of the force with depth ($\alpha\simeq1$) at small depth, followed by a saturation at larger depth ($\alpha\simeq0$) attributed to the Janssen screening wall effect, and finished by an exponential increase very close to the bottom wall.

\begin{figure}[b]
\centerline{\includegraphics[width=\linewidth]{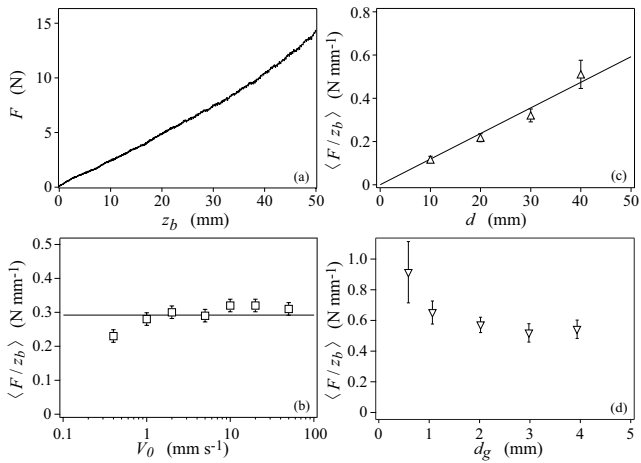}}
\caption{(a)~Drag force $F$ on the cylinder as a function of its penetration depth $z_b$ for $V_0=10$\,mm s$^{-1}$, $d = 20$\,mm, and $d_g=1$\,mm. Depth force variation $\langle F/z_b\rangle$, with $\langle\cdot\rangle$ standing for an average of $z_b$ in the range $d/2 \le z_b \le H-d$, as a function of (b)~the velocity $V_0$ (for $d=20$\,mm and $d_g=1$\,mm), (c)~the cylinder diameter $d$ (for $V_0=5$\,mm s$^{-1}$ and $d_g= 1$\,mm), and (d)~the grain size $d_g$ (for $V_0=5$\,mm s$^{-1}$ and $d =40$\,mm). Symbols are experimental data and the solid lines correspond to the best fits of the data with (b)~$\langle F/z_b\rangle\simeq 0.29$\,N mm$^{-1}$ and (c)~$\langle F/z_b\rangle/d \simeq 0.012$\,N~mm$^{-2}$.}
\label{Fig10}
\end{figure}

We present here force measurements associated with the velocity profiles presented above and corresponding to our geometry of a horizontal penetrating cylinder in vertical displacement in the granular packing. A typical force variation as a function of the penetration depth $z_b$ is shown in Fig.~\ref{Fig10}(a), where $z_b$ is the position of the cylinder bottom, with thus $F=0$ when $z_b=0$. Figure~\ref{Fig10}(a) shows that $F$ increases with $z_b$ whereas the grain velocity profile around the cylinder has been shown to be stationary in section~\ref{subsec:meanvelocity}. The $F(z_b)$ variation is about linear, and if one considers a hydrostatic equivalent ``pressure'' $p=\rho gz$ that varies linearly with depth in the granular material, this would lead to the Archimedean force $F_A=\rho g\pi bd^2/4$ equal to about 0.2\,N for the case of Fig.~\ref{Fig10}(a), which is thus negligible when compared to the measured drag force.

We do not observe any drastic change in the rate of increase of $F$ with $z_b$, which may be due to the fact that we do not explore very large depth values nor depth values close to the bottom wall. The mean slope of $F(z_b)$ can be extracted and is displayed in Fig.~\ref{Fig10} for different (b) penetration velocities $V_0$, (c) cylinder diameters $d$, and (d) grain diameters $d_g$. One can see that the drag force is approximately independent of $V_0$ and proportional to the cylinder diameter $d$. The $F$ variation with the grain diameter $d_g$ is more complex: When $F$ seems to be about constant for large grain diameters ($d_g \gtrsim 1$\,mm), it seems to increase with decreasing $d_g$ for small grain diameters ($d_g\lesssim 1$\,mm) as already mentioned by \cite{Albert99,Chehata03}.

\section{Discussion}
\label{sec:discussion}

\begin{figure}[b]
\centerline{\includegraphics[width=\linewidth]{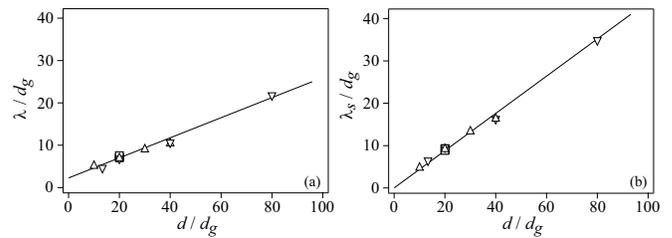}}
\caption{Dimensionless characteristic lengths of the velocity profiles, (a)~$\lambda/d_g$ and (b)~$\lambda_s/d_g$, as a function of the size ratio $d/d_g$. (---)~Best fit of the data (same symbols as in Fig.~\ref{Fig05}) with equations (a)~$\lambda=2.3d_g+0.24d$ and (b)~$\lambda_s=0.44d$.}
\label{Fig11}
\end{figure}

The velocity field and the drag force around a moving cylinder in granular matter show important differences compared to the case of a Newtonian fluid. As a matter of fact, the drag force in granular matter is independent of velocity in the explored range of velocities but increases with depth. This differs from Newtonian fluids for which the hydrodynamic drag increases with velocity but is independent of depth, as long as fluid viscosity is independent of pressure so that pressure only leads to a constant Archimedean buoyancy term. Concerning the velocity field, the perturbation created by the cylinder in granular matter appears localized near the cylinder with an exponential radial decrease ($B(r) \sim r[1-\exp(-r/\lambda)]$), which differs significantly from the Newtonian case in either the inviscid case for which the perturbation relaxes with a power law of the distance ($B(r) = r(1-d^2/4r^2)$) or the viscous case for which the perturbation relaxes slowly at large distance from the cylinder ($B(r)\simeq r \ln(d/2r)/\ln(\mathrm{Re})$ for $\mathrm{Re}\ll 1$ and $r\gg d$). The present granular case is closer to yield stress fluids where a strong localization is also observed \cite{Atapattu95,Dollet07}. All our $\lambda$ measurements at different velocities, and cylinder and grain diameters, collapse in Fig.~\ref{Fig11}(a) when made dimensionless with the grain size $d_g$ leading to a linear increase of $\lambda/d_g$ with the size ratio $d/d_g$, with a small but non zero value $2.3 \pm 0.5$ for vanishing $d/d_g$ and slope $0.24 \pm 0.03$. This means that the characteristic length $\lambda$ of mean velocity radial variation is mainly governed by the cylinder diameter $d$ as long as the size ratio $d/d_g$ is greater than about 10, which is always the case here in contrast to small intruder size situations studied, e.g., in \cite{Candelier09}. In contrast, Fig.~\ref{Fig11}(b) shows undoubtedly that the ``slipping length" $\lambda_s$ depends only on the cylinder diameter $d$ with the scaling $\lambda_s = (0.44 \pm 0.4)d$ very close to $d/2$.

\begin{figure}
\centerline{\includegraphics[width=0.7\linewidth]{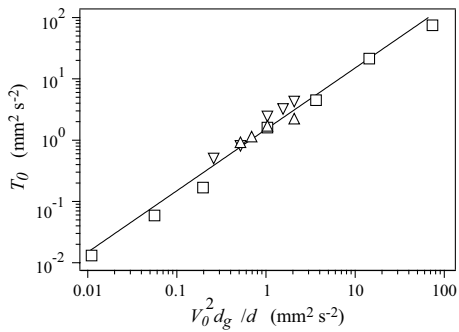}}
\caption{Temperature plateau $T_0$ as a function of $V_0^2d_g/d$. (---)~Fit of equation $T_0 = 1.5 V_0^2d_g/d$ through the data (same symbols as in Fig.~\ref{Fig08}).}
\label{Fig12}
\end{figure}

\begin{figure}
\centerline{\includegraphics[width=\linewidth]{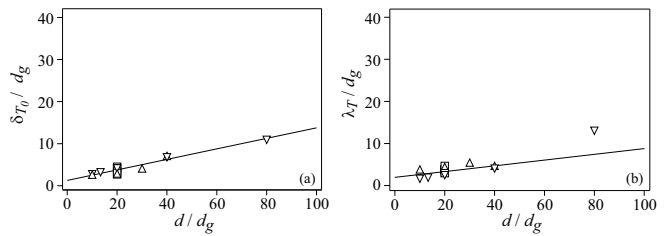}}
\caption{Dimensionless characteristic length (a)~of temperature plateau $\delta_{T_0}/d_g$ and (b)~of temperature exponential decrease $\lambda_T/d_g$ as a function of the size ratio $d/d_g$. (---)~Fit of equation (a) $\delta_{T_0} = 1.4 d_g + 0.12 d$ and (b) $\lambda_T= 2 d_g +0.07 d$ through the data (same symbols as in Figs.~\ref{Fig08} and \ref{Fig09}).}
\label{Fig13}
\end{figure}

As the typical granular temperature $T_0$ of the granular flow around the cylinder is connected with the relative velocity $V_0$, there is clearly a strong coupling between the mean velocity profile and the temperature profile.
In that sense, the present situation resembles the case of the motion of a hot cylinder or sphere in a fluid of temperature dependent viscosity. This last situation has already been studied as it corresponds to the geophysical situation of the ascending motion of hot diapirs \cite{Morris82,Ansari85}. In this last case, the flow is also strongly localized close to the hot sphere where the viscosity is much lower than far away, as heat diffuses into a fluid layer near the object, thus lowering the fluid viscosity. Indeed, stress continuity causes regions of low viscosity to be regions of large strain rates, producing a temperature-induced shear band. In this last case, heat is produced outside the flowing material by the hot sphere.
In the present granular case, the ``heat" production is due to the flow itself, and the scaling of the temperature plateau $T_0$ appears very close to $T_0 = (1.5 \pm 0.5) V_0^2 d_g/d$ in the log-log plot of Fig.~\ref{Fig12} where all the data for different $V_0$, $d$, and $d_g$ collapse along a slope 1. We have seen in the preceding section (Fig.~\ref{Fig09}) that the radial extension $\delta_{T_0}$ of the temperature plateau close to the cylinder is linked to the characteristic length $\lambda$ of the radial variations of the mean velocity. The plot of all data of $\delta_{T_0}/d_g$ at different $V_0$, $d$ and $d_g$ as a function of $d/d_g$ in Fig.~\ref{Fig13}(a), exhibits a small but non zero value $1.4 \pm 0.2 $ for vanishing $d/d_g$ and a slope $0.12 \pm 0.02$ corresponding roughly to $\delta_{T_0} \sim \lambda/2$. As $\lambda$, $\delta_{T_0}$ is mainly governed by the cylinder diameter $d$ as long as the size ratio $d/d_g$ is larger than about 10. All these findings mean that the granular heat is created by the shear flow $V_0/\lambda \sim V_0/d$ close to the cylinder, in a region of radial extension $\sim \lambda$ (more precisely $\delta_{T_0} \sim \lambda/2$).
This shear flow leads to velocity fluctuations $V_0 d_g/\lambda \sim V_0 d_g/d$ at the grain scale that in turn produce the granular temperature $(V_0 d_g/d)^2$. The observed scaling for the temperature plateau $T_0 \sim V_0^2 d_g/d$ should come from an equilibrium between the corresponding ``heat" production by the high shear close to the cylinder and the granular dissipation away from the cylinder. The characteristic length $\lambda_T$ of the exponential decrease of the temperature observed far away from the cylinder is mainly governed by the grain size $d_g$: all the data of $\lambda_T/d_g$ at different $V_0$, $d$ and $d_g$ plotted in Fig.~\ref{Fig13}(b) as a function of the size ratio $d/d_g$ only show a slight increase with $d/d_g$ of slope $0.07 \pm 0.07$ from the value $2 \pm 1$ at vanishing $d/d_g = 0$.

The flow localization, observed in our case in the vicinity of the cylinder, is indeed reported in various physical systems.
It is a common feature of matter with granularity as recently reviewed by \cite{Schall10}. Such shear bands are currently observed for flow of granular materials, foams, or emulsions. In such systems, the thicknesses of the shear bands are usually of few grains or bubbles. Despite the generality of such observations, there is still no clear unifying framework for such shear bands \cite{Schall10}. In the case of yield stress fluids, the shear-bands widths may be large compared to the microscopic sizes \cite{Ovarlez09}, and shear localization is observed in Newtonian fluid flow close to walls at temperatures larger than the temperature of the fluid far away \cite{Morris82,Ansari85}.

Note that in a cylindric Couette device, Refs.~\cite{Losert00,Bocquet01} have reported a velocity profile for the sheared dry grains independent of the cylinder angular velocity and also independent of the ``pressure" imposed by a possible external upward or downward air flow. This has to be related to our case in which the velocity profile around the moving cylinder does not depend on the depth and thus on the effective ``pressure''. References \cite{Losert00,Bocquet01} also reported a small zone of constant granular temperature $T_0 \sim V_0^2$ and extension $\delta_{T_0} \simeq 2.8 d_g$ close to the rotating cylinder followed by an exponential decrease at larger distance with a characteristic length $\lambda_T \simeq 4.7d_g$. The temperature profile we find in the present study for the flow around a cylinder is similar, with a similar scaling of $\lambda_T$ but different scalings for $T_0$ and $\delta_{T_0}$ that should come from geometric consideration.

In \cite{Seguin11}, a model was briefly presented for the granular flow around the cylinder based on continuous conservation equations for mass, momentum, and granular temperature. The phenomenological equations relating normal and shear stresses, density, and temperature are given by the kinetic theory of granular gases \cite{Jenkins83} in the dense limit. In this approach, the granular material may be viewed as a fluid whose viscosity decreases, at a given pressure, with the granular temperature. From a thermal point of view, the heat is produced by shearing the fluid, then diffuses into the material, and is dissipated during particles collisions. It follows from such an approach that a sheared zone increases the granular temperature, which fluidizes the material, increasing the ability to flow. The numerical simulation of mechanical and heat equations shows indeed the presence of such a fluidified zone near the obstacle \cite{Seguin11}. The constitutive equation relating the viscosity to the temperature and the pressure leads to a velocity-independent drag force in the creeping flow regime, as observed experimentally.

\begin{figure}
\centerline{\includegraphics[width=0.7\linewidth]{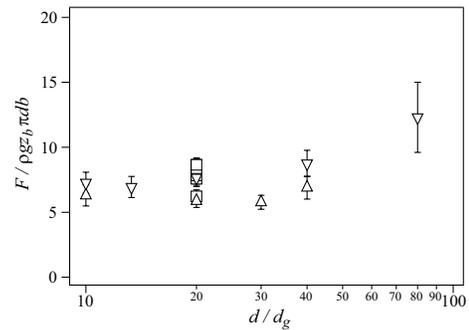}}
\caption{Normalized drag force $F/(\rho gz_b\pi db)$ on the cylinder as a function of the size ratio $d/d_g$. Same symbols as in Fig.~\ref{Fig10}.}
\label{Fig14}
\end{figure}

Considering now our observations for the resulting drag force on the cylinder, one can infer an effective friction coefficient $\mu = F/(\rho gz_b\pi db)$ based on a hypothetical effective hydrostatic pressure $\rho gz$ at the depth $z$ inside the granular material. All the data from different $V_0$, $d$, and $d_g$ collapse when plotted as a function of the size ratio $d/d_g$ in Fig.~\ref{Fig14}, meaning that no other finite size effect (such as with the container sizes $H$, $L$, or $b$) is observed in the present experiment. The effective friction coefficient appears to be greater than 5 with a plateau value of about $7 \pm 2$ at low enough $d/d_g$ ($d/d_g\lesssim 50$) and a possible increase for larger $d/d_g$ values ($d/d_g \gtrsim 50$). This unusual value of the friction coefficient, which is about 20 times higher than the usual values of friction coefficients in dense granular flows \cite{GDRMidi04}, has already been mentioned in \cite{Katsuragi07} for the depth-dependent term of the drag force extracted from impact experiments. This means that the ``pressure" is undoubtedly far from hydrostatic in the present situation even if the force variation with depth is almost linear.

\section{Conclusions}
In this paper, we have studied both velocity profile and force measurements for the granular flow around a horizontal cylinder in the vertical penetration case. While the force increases with the depth, the velocity profile is shown to be stationary during the penetration with a well-defined average and some erratic fluctuations. The granular flow being close to bidimensional and incompressible, the stream function has been extracted from the velocity measurements. The flow perturbation is strongly localized close to the cylinder, exhibiting an exponential radial decrease away from the cylinder with a characteristic length $\lambda$ that scales mainly with the cylinder diameter $d$ for large enough cylinder to grain size ratio $d/d_g$: $\lambda \simeq 2 d_g + d/4 \approx d/4$ for $d/d_g \gtrsim 10$. The velocity fluctuations quantified by the so-called granular temperature $T$ show also a strong localization near the penetrating cylinder, with a plateau value $T_0$ in a narrow crown of extension $\delta_{T_0}$ around the cylinder followed by an exponential radial decrease with a small characteristic length $\lambda_T$. The scaling of $T_0$ appears to be simply $T_0 \sim V_0^2 d_g/d$ coming from a balance between the ``granular heat" production by the shear flow $V_0/\lambda$ ($\sim V_0/d$ for $d/d_g \gtrsim 10$) over the distance $\delta_{T_0} \sim \lambda/2$ ($\sim d/8$ for $d/d_g \gtrsim 10$) and the granular dissipation far away ($r > \delta_{T_0}$). This granular dissipation which is at the grain scale leads to a characteristic length scaling $\lambda_T$ for the temperature decrease of a few grain diameters: $\lambda_T \sim 3d_g$.

In conclusion, the granular flow around a moving cylinder is very different from the flow of a Newtonian fluid either in a viscous or inertial regime: For the Newtonian fluid case, whatever the regime is, the drag force increases with the velocity and does not depend on the depth, whereas the contrary is observed for the granular case for which the drag force does not depend on the velocity but increases with the depth. In an upcoming paper, we will detail a possible hydrodynamic modeling of this non classical fluid behavior based on kinetic theory adapted for dense dissipative granular systems as briefly presented in \cite{Seguin11}. The shear localization characterized in the present experiment is certainly important to understand the influence of a close wall on the motion of objects in some practical situations such as the dynamics of impacting spheres \cite{Seguin08,Durian08}. In addition, other experiments have to be performed to understand the very different drag force experienced by an object when moving down or up in penetration or withdrawal situations, as already observed by \cite{Hill05} and \cite{Schroter07}.

We are grateful to A. Aubertin and R. Pidoux for the development of the experimental setup, and E.~J. Hinch and O. Pouliquen for stimulating discussions. This work is supported by the ANR project STABINGRAM No. 2010-BLAN-0927-01.

\bibliography{Seguin}
\end{document}